\documentclass[twocolumn,aps,showpacs]{revtex4}
\input epsf
\begin{document}
\title{Noise dressing of the correlation matrix 
of factor models}
\author{F. Lillo$^{*}$ and R. N. Mantegna$^{\dag}$\ }
\affiliation{$^{*}$Santa Fe Institute, 
1399 Hyde Park Road, Santa Fe NM 87501\\
$^{\dag}$ INFM and Dipartimento di Fisica e Tecnologie Relative, 
Universit\`a di Palermo, Viale delle Scienze, I-90128 Palermo, Italia.
}

\begin{abstract}
We study the spectral density of factor models of 
multivariate time series. By making use of the Random Matrix
Theory we analytically quantify the effect of noise dressing 
on the spectral density due to the finiteness of the sample.
We consider a broad range of models ranging from one factor
models in time and frequency domain to hierarchical multifactor 
models.  
\end{abstract} 
\maketitle

The extraction of information from a multivariate time series is
a central issue in many applications. Several methods has been
introduced to this end, ranging from principal component analysis
to clustering methods \cite{mardia,cluster}. 
The simplest and more
widespread models of multivariate time series are factor models.
In these models the dynamics of each variable is the linear combination
of a given number of factors plus an idiosyncratic noise term.
The coefficients of the linear combination and the intensity of the
noise terms are specific of each variable and assumed for simplicity 
to be time independent. Examples of such models are Capital Asset Pricing
Model or CAPM (one factor) and Arbitrage Pricing Theory (multifactor)
in the financial domain \cite{campbell}. 
Another class of factor models describes
the dynamics of the variables driven by one or more 
sinusoidal signals of given frequency with each variable  
characterized by a different phase. This kind of models 
as been recently applied to gene expression analysis 
during cell cycle
obtained by microarray data \cite{maritan,brown}. In this second
case the variables are following the common factor(s) in
frequency domain rather than in real time.
The more general multifactor model for $N$ variables $x_i(t)$ 
($i=1,...,N$) can be written as
\begin{equation}
x_i(t)=\sum_{j=1}^{K}\gamma_i^{(j)} f_j(t)+\gamma_i^{(0)}\epsilon_i(t).
\end{equation}
In this equation $K$ is the number of factors $f_j(t)$,
$\gamma_i^{(j)}$ is a constant describing the weight of factor $j$
in explaining the dynamics of the variable $x_i$, and $\epsilon_i(t)$ 
is a Gaussian zero mean noise term with unit variance. In Eq.(1)
we assume that the factors are uncorrelated one with each other,
i.e. $\langle f_i(t)f_j(t)\rangle=\delta_{ij}$, where the symbol
$\langle ... \rangle$ indicates an average in time. Also the noise
terms are uncorrelated one with each other and with the factors,
i.e. $\langle \epsilon_i(t)\epsilon_j(t)\rangle=\delta_{ij}$ and 
$\langle f_i(t)\epsilon_j(t)\rangle=0$. Since in the rest
of this paper we are interested in studying the linear correlation 
coefficients, without loss of generality we assume that
all the variables $x_i$ have zero mean and unit variance.
These assumptions
fix the value $(\gamma_i^{(0)})^2=1-\sum_j(\gamma_i^{(j)})^2$.

Multivariate methods 
are designed to extract the information
on the number of factors and on the composition of the groups.
On the other hand any real experiment is performed on a finite
sample of $T$ records for each variable and the measured 
quantities in the analysis are unavoidably dressed by noise.
In this letter
we study the role
of the noise in dressing the properties of the spectral density of
the correlation matrix. The correlation matrix $\mbox{{\bf C}}$ 
is the $N\times N$
symmetric matrix whose entries are the linear correlation coefficient
between each pair of variable $x_i$ and $x_j$. The object of
our study is the spectral density of the sample correlation
matrix ${\bf \hat C}$. The square root of the eigenvalues 
of ${\bf \hat C}$ are called singular values.
We will make use of the Random Matrix Theory \cite{metha}
to compute the role of noise dressing on the spectrum of
correlation matrices of factor models. Most of the results
we derive are valid in the limit of $N \to \infty$ and
$T \to \infty$, even though for real large matrices 
the approximation is quite good.
The application of Random Matrix Theory to the noise dressing
of correlation matrices has been recently addressed 
\cite{denby,sengupta,sosh}
and applied to the study of financial correlation matrices 
\cite{bouchaud,stanley,sornette}. In Ref. \cite{bouchaud,stanley} 
the null hypothesis
used to compare real sample correlation matrices is a model
of uncorrelated variables, or in our language a zero factor model
or random model. In the random model each variable is 
described only by a random Gaussian variable $\epsilon_i(t)$.
The corresponding class of random matrices is called Wishart
matrix in statistical literature.
The correlation matrix is the identity
matrix and the spectral density is $\rho(\lambda)=N\delta(\lambda-1)$.
The noise dressing of the spectrum of the sample correlation matrix
has been derived in \cite{denby,sengupta}.
In the limit $T,N \to \infty$, with a fixed
ratio $Q=T/N \geq 1$, the spectral density of the correlation matrix
is given by
\begin{equation}
\rho(\lambda)=\frac{T}{2\pi\lambda}\sqrt{(\lambda_{max}-\lambda)
(\lambda-\lambda_{min})}, 
\end{equation}
where $\lambda_{min}^{max}=(1+1/Q\pm 2\sqrt{1/Q})$. Since the procedure
used to obtain this result is very similar to the one we use below, 
here we summarize it briefly.
One introduces the resolvent
\begin{equation}
{\cal G}(z)=\sum_{n=1}^N\frac{1}{z-\lambda_n},
\end{equation}
which is related to the spectral density through
\begin{equation}
\rho(\lambda)=\frac{1}{\pi}\lim_{\epsilon\to 0}
Im[{\cal G}(\lambda -i\epsilon)].
\end{equation}
The resolvent is equal to ${\cal G}(z)=\partial_z \ln \det(z-\rho)$.
By making use of the replica trick and by performing a saddle point
approximation one finds an equation for the ensemble average of the
resolvent $G(z)=\langle {\cal G}(z)\rangle_{ens}$ and through 
Eq. (4) the spectral density is obtained. The symbol
$\langle ... \rangle_{ens}$ indicates an average on the ensemble 
of variables.

Even if the random model can be sometimes a starting null hypothesis,
a factor model should be used as null hypothesis when one 
suspects the presence of common factors in the dynamics of
an ensemble of variables. As a first example we shall consider 
the one factor model in which
the dynamics of each variable is controlled by a single common factor.
The equations describing the one factor model is given by Eq. (1)
with $K=1$.
The parameter $\gamma_i^2$ gives the 
fraction of variance explained by the common factor $f(t)$. 
 It is direct to show that the correlation
coefficient between variable $i$ and $j$ is $\gamma_i \gamma_j$.
The correlation matrix of the one factor model can therefore be written as
${\bf C}={\bf A}+{\bf bb^+}$, where ${\bf A}=diag(1-\gamma_i^2)$ is a diagonal $N \times N$
matrix and ${\bf b^+}=(\gamma_1,...,\gamma_N)$ is a row vector. The characteristic
equation of ${\bf C}$ can be calculated by using the Sherman-Morrison 
formula \cite{numrec}, and the result is
\begin{equation}
\det({\bf C}-{\bf I_N}\lambda)=\prod_{i=1}^N(1-\gamma_i^2-\lambda)
\left[1+\sum_{i=1}^N\frac{\gamma_i^2}{1-\gamma_i^2-\lambda}\right]=0.
\end{equation}
In the case of a degenerate one factor model, i.e. when 
$\gamma_i=\gamma$ for all values of $i$, the characteristic equation
can be solved and the spectrum is composed by 
a large eigenvalue $\lambda_1=1+(N-1)\gamma^2\simeq N\gamma^2$ and
$N-1$ degenerates eigenvalues $\lambda_i=1-\gamma^2$ where $i=2,...,N$
\cite{sornette}.
We are not able to solve Eq.(5) in the non degenerate case, but
we are able to 
provide an approximate form of the spectrum when $N$ is large. 
In the non-degenerate
case we can still expect a spectral density composed by a large
eigenvalue (high part) and $N-1$ small eigenvalues (low part). 
The large eigenvalue
can be obtained by putting to zero the term in square brackets in
Eq.(5). The largest eigenvalue is much larger than $1-\gamma_i^2$
for any $i$ and we can therefore approximate the characteristic equation
as $\sum \gamma_i^2/\lambda_1\simeq1$, i.e.
$\lambda_1\simeq\sum_{i=1}^N\gamma_i^2=N\langle\gamma_i^2\rangle_{ens}$.
In order to have an insight on the spectral density for the other
$N-1$ eigenvalues we assume that the $\gamma_i$ are distributed
according to a given probability density $P(\gamma_i)$. We make
the ansatz that the distribution of $\lambda_i$ is given by
$P(\lambda)=P(\gamma_i)d\gamma_i/d\lambda$ where 
$\gamma_i=\sqrt{1-\lambda}$. The idea behind this ansatz is
that the relation between eigenvalues and $\gamma_i$ is 
the same as in the degenerate case. For example if $\gamma_i$
is distributed uniformly in a subinterval $[m-d,m+d]$ of $[0,1]$,
the distribution of the $N-1$ eigenvalues is given by
$P(\lambda)=(4d)^{-1} (1-\lambda)^{-1/2}$. 
Note that under our ansatz this part of the spectrum is bounded
from above by the value $\lambda=1$. In panel (a) of Fig. 1 we 
show the low part of the spectral density
of a one factor model describing $N=2000$ variables.
The $\gamma_i$ are distributed exponentially with $\bar \gamma=0.25$.
The line is the theoretical prediction based on our ansatz. 
The agreement between data and the ansatz is quite good.

\begin{figure}[t]
\epsfxsize=8.5cm
\centerline{\epsfbox{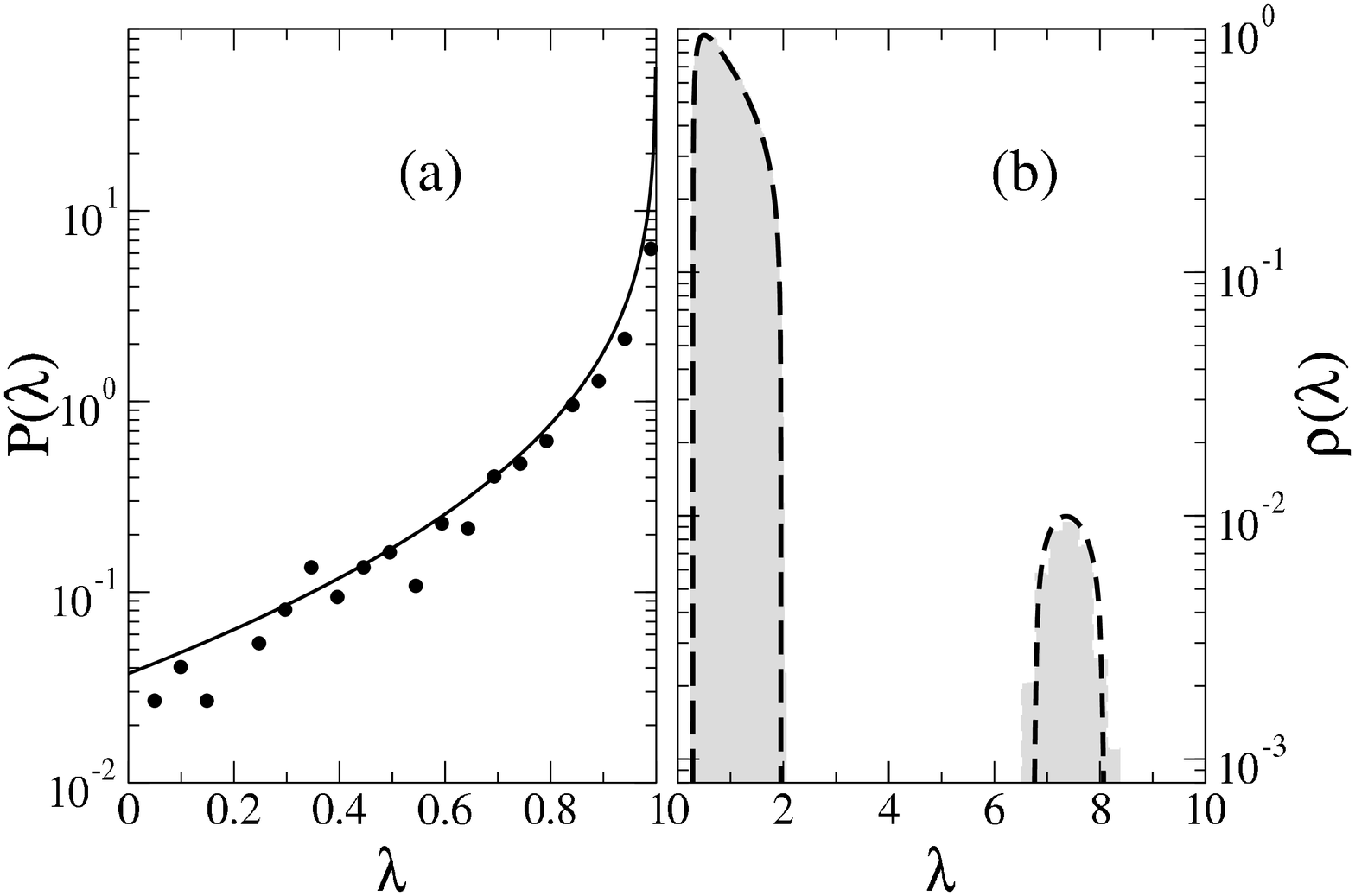}}
\caption{(a) The low part of the spectrum of a single realization 
of a non degenerate 
one factor model (circle). In this case $N=2000$ and $\bar \gamma=0.25$. 
The continuous
line is the prediction based on the ansatz discussed in the text. 
(b) Spectral density of a degenerate one factor model. The gray
areas are the average over $1000$ numerical simulations of a one factor
model of $N=100$ variables for $T=500$ time steps. The value of $\gamma$
is $0.25$. The dashed line is the theoretical prediction. }
\label{fig1}
\end{figure}

We now consider the noise dressing of the spectrum due to the 
finiteness of the sample. In other words we assume that our data
are described by a one factor model (degenerate or not) and we
assume that we have $T$ synchronous records for each $x_i$ variable.
By using the arguments of Ref. \cite{sengupta} we can prove that 
the ensemble averaged resolvent of a one factor model of 
$N$ variables for $T$ time steps is determined by the equation
\begin{equation}
G(z)=\frac{T}{z-\sum_{i=1}^{N}\frac{\lambda_i}{T-\lambda_i G(z)}}.
\end{equation}
For the degenerate one factor model discussed above the equation
for the resolvent is a third degree algebraic equation \cite{supplement}, which
can be solved exactly. The spectral density can be obtained analytically
from Eq.s (4) and (6), even if the expression is quite long. As expected
the spectral density is different from zero in two intervals, 
one for the $N-1$ small eigenvalues and one for the large 
eigenvalue $\lambda_1$. 
Numerical calculations and analytical
considerations show that the low part of the spectrum is well
fitted by a functional form of Eq.(2). Moreover 
the width of the two intervals 
scale with the parameter of the model as
$ \Delta_0\sim(1-\gamma^2)\sqrt{N/T}$ and 
$\Delta_1\sim N\gamma^2/\sqrt{T}$,
where $\Delta_0$ ($\Delta_1$) is the width of the low (high)
part of the spectrum. Panel (b) of Fig. 1 shows the comparison 
of theoretical prediction
and numerical simulations of a degenerate one factor model. 
The agreement is very good in 
the whole range of eigenvalues. It is worth noting that such an
agreement is obtained also when $T<N$.

\begin{figure}[t]
\epsfxsize=8.5cm
\centerline{\epsfbox{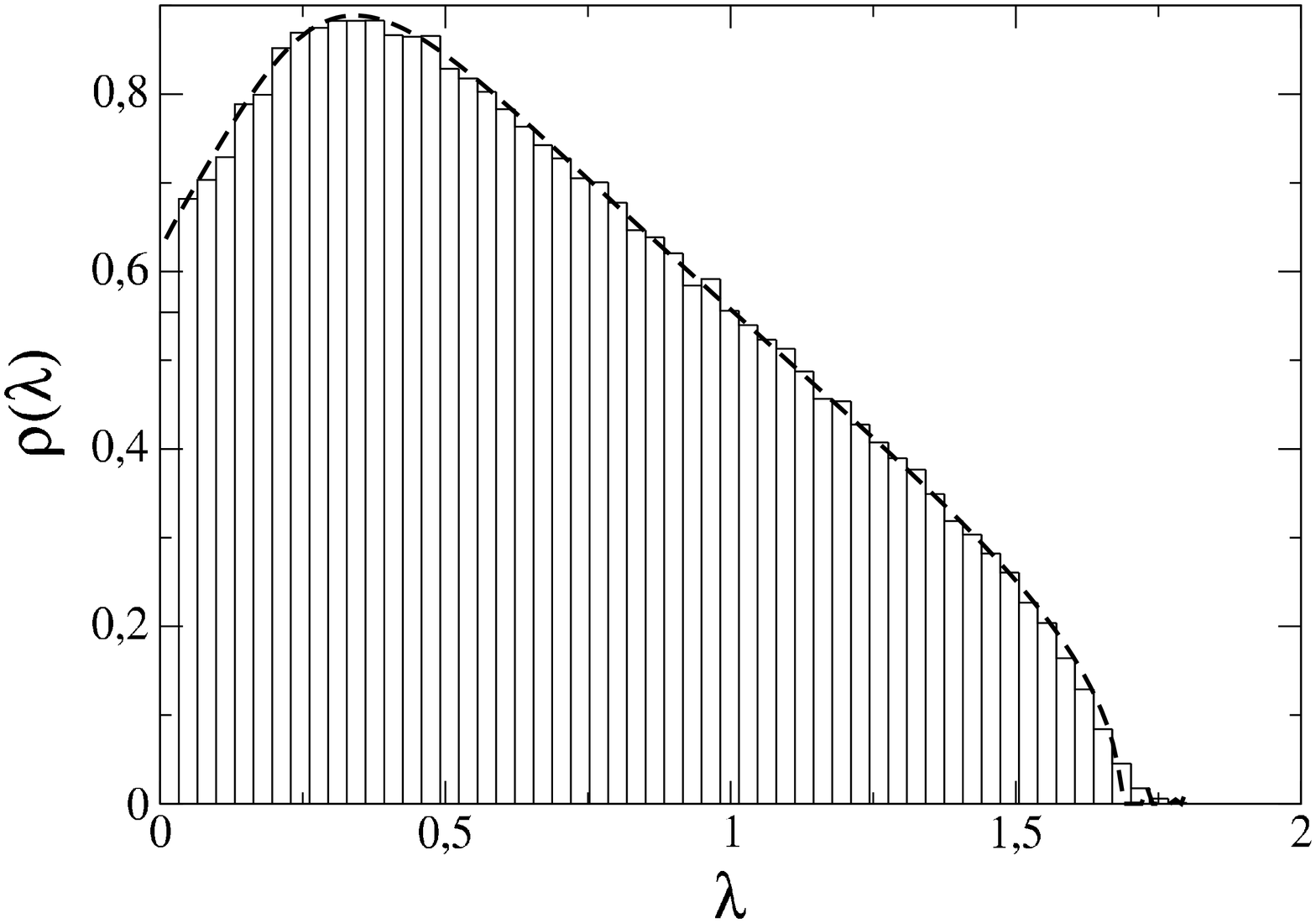}}
\caption{Low part of the spectral density of a completely non degenerate 
one factor model, i.e. a one factor model in which $\gamma_i$ is distributed
uniformly between 0 and 1. The dashed line is the theoretical result
obtained through the theory developed in the text.}
\label{fig2}
\end{figure}

We come now to the more complicated case of non degenerate 
one factor model. In order to find the spectrum one should solve
Eq.(5) for all the eigenvalues and then solve the $(N+1)$th degree
polynomial of Eq.(6). This task is too complicated even numerically.
We use a different approach making use of the fact that $N$ is large.
The sum in the denominator of Eq. (6) is split in a term for 
$\lambda_1$ plus a sum over the remaining $N-1$ small eigenvalues.
This last term can be computed as $N-1$ times the average of 
$\lambda/(T-\lambda G(z))$ over $P(\lambda)$ introduced in our ansatz.
In other words, Eq. (6) for the resolvent becomes
\begin{equation}
G(z)=\frac{T}{z-\frac{\lambda_1}{T-\lambda_1 G(z)}-
(N-1)\langle \frac{\lambda_i}{T-\lambda_i G(z)}\rangle_{ens}},
\end{equation}
where $\lambda_1\simeq 1+(N-1)\langle \gamma_i^2
\rangle_{ens}$. In general the average term in (7) is not a rational
function and therefore Eq.(7) cannot be reduced
to an algebraic equation in $G$. In order to solve the
complex transcendental equation (7) we introduce a simple algorithm.
The average term in (7) depends typically
on the dispersion of the $P(\gamma_i)$ at the second order. 
Therefore the low part of the spectrum of a non degenerate
one factor model with a small dispersion in $\gamma_i$ is not
very different from a degenerate one-factor model with
$\gamma_{eff}=\sqrt{\langle \gamma_i^2\rangle_{ens}}$. We can 
therefore use the value of the resolvent of this effective
degenerate one factor model as the starting point for the searching
of the solution of Eq. (7). Quite surprisingly this method
works also when the dispersion of the $\gamma_i$ is high.
For example for a uniform distribution of $\gamma_i$,
the average term in Eq. (7) involves two inverse hyperbolic
tangent functions \cite{supplement}. In Fig. 2 we show the low part 
of the spectrum 
for a completely degenerate one factor model, i.e. a model in
which $\gamma_i$ is uniformly distributed between $0$ and $1$.
We see that the agreement between the theory and the simulations 
is very good. A similar good agreement is observed for 
exponentially distributed $\gamma_i$ \cite{supplement}.
It is worth noting that in the general case 
of a degenerate one factor model the low part of the spectrum
is not compatible with a Wishart form of Eq.(2).  
  
The results obtained for the one factor model can be easily 
extended to multifactor model. When the factors are stochastic
and uncorrelated one with each other the structure of the correlation
matrix is given by the composition of the groups of variables
correspondent to the factors. The simplest case is when each
variable belongs to one and only one groups, i.e. its dynamics
is determined by only one factor and by the idiosyncratic noise.
In this case the correlation matrix is block diagonal. 
The correlation coefficient between variables belonging to 
different groups is zero, while, when the variables $i$ and $j$
belong to the same group $k$, the correlation coefficient is 
$\gamma_i^{(k)}\gamma_j^{(k)}$. The spectral density of 
this kind of models is simply given by the superposition of the 
spectral densities of $K$ one factor models. The theory of
noise dressing follows directly from Eq. (6) in which the 
number of distinct eigenvalues is $2K$. The equation for $G(z)$
is therefore an algebraic equation of degree $2K+1$.
Again if the distributions of $\gamma_i^{(k)}$ are given, one
can solve the non degenerate case by using the same arguments
of the one factor model. Clearly the computational task increases 
with the number of factors.

An interesting generalization of multifactor models occurs when there is 
a hierarchical overlap between different groups. 
To give a concrete example, let us
consider a portfolio of stocks. As a first approximation 
we can consider the portfolio as composed by a large group 
following a common factor (the market factor in CAPM) and a certain
number of groups homogeneous in economic activity following
a sectoral factor, such as, for example, the oil companies
or the technological group. In this case the composition of the groups
induces a hierarchical structure to the correlation matrix. 
Also this kind of models can be solved.
 We present here a simple example
in which the $N$ variables follow a common factor with a 
constant $\Gamma$. Moreover the set of variables is divided
in two groups. We have $n_1$ variables following
the first subfactor with constant $\gamma_1$ and $n_2=N-n_1$ 
variables following the second subfactor with constant 
$\gamma_2$. 
The correlation matrix
of this model is a block matrix composed
by a block $n_1\times n_1$ matrix whose entries are 
$\Gamma^2+\gamma_1^2$ and $1$ on the diagonal and a block
$n_2\times n_2$ matrix whose entries are 
$\Gamma^2+\gamma_2^2$ and $1$ on the diagonal. 
The elements in the out diagonal block submatrices are equal 
to $\Gamma^2$.

The spectrum of this matrix is composed  
by two large eigenvalues given by
\begin{eqnarray}
\lambda_{\pm}=\frac{1}{2}(2+\gamma_1^2(n_1-1)+\gamma_2^2(n_2-1)+\Gamma^2(n_1+n_2-2)
 \nonumber \\ \pm  
\sqrt{A^2+
\Gamma^4(n_1+n_2)^2+2A\Gamma^2(n_1-n_2)}),&
\end{eqnarray}     
where $A=(\gamma_1^2(n_1-1)-\gamma_2^2(n_2-1))$
and $n_1-1$ eigenvalues equal to $1-\Gamma^2-\gamma_1^2$ and 
$n_2-1$ eigenvalues equal to $1-\Gamma^2-\gamma_2^2$.
Again by making use of Eq. (6) it is possible to find the 
effect of noise dressing by solving the corresponding 5th degree
algebraic equation \cite{supplement}. 
To give the general idea behind the derivation of result (8)
and of its generalization to more complex hierarchical models,
we note that correlation matrix of degenerate hierarchical
factor models can be written in terms of the identity matrix
and unit matrices ${\bf J}_{mn}$ (i.e. a $m \times n$ matrix
consisting of all 1s). These matrices form a closed algebra 
under multiplication, for example ${\bf J}_{nm}{\bf J}_{mp}=m{\bf J}_{np}$.
The determinant leading to the characteristic equation is directly obtained 
by using the block submatrices in which the correlation matrix
can be partitioned. The above mentioned algebraic properties 
allows to reduce the characteristic equation in terms of 
$\det(a{\bf I}_n+b{\bf J}_{nn})=a^{n-1}(a+bn)$.

The last class of model we consider is given by a one factor
model in which the synchronization with the factor is in frequency
rather than in time domain. Specifically, let us consider a
factor model in which the dynamics of the variables is described by 
the equation
\begin{equation}
x_i(t)=\gamma\sqrt{2}\sin(\omega t+\phi_i)+\gamma^{(0)}\epsilon_i(t).  
\end{equation}
The ensemble average of the product of two variables in two distinct
instants of time can be written
as $\langle x_i(t) x_j(t')\rangle_{ens}=C_{ij}D_{tt'}$,
where $C_{ij}=\delta_{ij}$ and 
$D_{tt'}=(\gamma^2\cos(\omega(t-t'))+(1-\gamma^2)\delta_{tt'})$.
The model described in Eq. (9) is the first 
approximation of the dynamics of the level of expression
of genes during cell cycle as detected in microarray experiments
\cite{maritan,brown}. In this case the frequency $\omega$ is
related to the duration of the cell cycle. Microarray experiments
have usually a very small number of time points compared with
the number of variables (genes). This fact leads to an heavy 
dressing of the correlation matrix by noise, and hence a 
careful characterization of the noise dressing is even more
important in this case.
In order to find the effect of noise dressing by using Random 
Matrix Theory we need to find the spectral density of the 
matrices ${\bf C}=(C_{ij})$ and ${\bf D}=(D_{tt'})$ \cite{sengupta}. 
In this case ${\bf C}$ is the identity matrix.
The spectrum of ${\bf D}$ can be found and it consists of two large
eigenvalues 
\begin{equation}
d_{1,2}=\frac{\gamma^2}{2}\left(T\pm\frac{\sin \omega T}{\sin \omega}
\right)+(1-\gamma^2),
\end{equation}
and $T-2$ eigenvalues equal to $d_i=(1-\gamma^2)$, where $i=3,...,T$. 
The equation for the resolvent is in this case (cfr. Eq.s (21-23) of Ref. 
\cite{sengupta})
\begin{equation}
G(z)=\sum_{i=1}^T \frac{1}{z-Q(z)d_i},
\end{equation}
where $Q(z)$ is the solution of the equations
\begin{eqnarray}
Q(z)=\frac{N}{T}\frac{1}{1-R(z)}, ~~~~~ 
R(z)=\frac{1}{T}\sum_{i=1}^T \frac{d_i}{z-d_iQ(z)}. 
\end{eqnarray}
The equation for $G(z)$ is a 4th degree algebraic equation that
can be solved exactly \cite{supplement}. 

In conclusion we have shown that the application of Random Matrix 
Theory allows to solve analytically the problem of the 
noise dressing of the spectral density of the correlation matrix
of a large class of factor models. 
This class includes one factor model in time and frequency domain
and hierarchical and non hierarchical multifactor models.
The main idea of the approach we are proposing can be summarized
as follows: (i) find the spectrum of the degenerate factor model;
this usually accounts to find the few distinct eigenvalues 
characterizing the spectrum. (ii) Find the effect of the noise
dressing in the degenerate case by using Eq. (6); this implies
the finding of roots of a low degree polynomial. (iii) Try to solve
the non degenerate case by averaging over the distribution of $\gamma$
parameters on the same lines of what has been done here for 
one factor model (see Eq.(7)). 
Our results can be applied in many different disciplines including
economics, finance, molecular biology and in general in any 
study in which factor models can constitute a good starting point
for modeling the simultaneous dynamics of many variables.

\end{document}